\documentclass[%
aps,prd,%
 amsmath,amssymb,
 reprint,%
 superscriptaddress,
nofootinbib,
floatfix,
preprintnumbers
]{revtex4-1}

\usepackage{graphicx}% Include figure files
\usepackage{dcolumn}% Align table columns on decimal point
\usepackage{bm}% bold math
\usepackage{xcolor}
\graphicspath{{Images/}}

\usepackage[caption=false]{subfig}
\usepackage[normalem]{ulem}
\usepackage{braket}
\usepackage{esdiff}
\usepackage{hyperref}
\usepackage{appendix}
\usepackage{mathtools}
\usepackage{amsmath}
\usepackage{calc}
\usepackage{float}

\begin{document}

% \interfootnotelinepenalty=10000

% \preprint{AAPM/123-QED}
\preprint{SLAC-PUB-17621, N3AS-21-014}

\title{Spectral splits and entanglement entropy in collective neutrino oscillations}
%Beyond the mean-field in collective neutrino oscillations: spectral splits and entanglement entropy}% Force line breaks with \\

\author{Amol V.\ Patwardhan}
\email{apatward@slac.stanford.edu} 
\affiliation{%
SLAC National Accelerator Laboratory,
Menlo Park, CA 94025, USA
}%
\author{Michael J.\ Cervia}
\email{cervia@gwu.edu}
\affiliation{%
Department of Physics, University of Wisconsin--Madison,
Madison, Wisconsin 53706, USA
}%
\affiliation{Department of Physics, The George Washington University,
Washington, District of Columbia 20052, USA
}
\affiliation{Department of Physics, University of Maryland, 
College Park, Maryland 20742, USA
}
% \affiliation{
% Department of Physics, University of California, Berkeley, CA 94720-6300, USA} 
\author{A.\ B.\ Balantekin}
\email{baha@physics.wisc.edu}
\affiliation{%
Department of Physics, University of Wisconsin--Madison,
Madison, Wisconsin 53706, USA
}%
% \author{S.\ N.\ Coppersmith}
% \email{snc@physics.wisc.edu}
% \affiliation{%
% Department of Physics, University of Wisconsin--Madison,
% Madison, Wisconsin 53706, USA
% }%
% \affiliation{%
% School of Physics, The University of New South Wales,
% Sydney, New South Wales 2052,  Australia}
% \author{Calvin W.\ Johnson}
% \email{cjohnson@sdsu.edu}
% \affiliation{%
%  Department of Physics, San Diego State University, San Diego,
%  California 92182-1233, USA
% }

\date{\today}% It is always \today, today,

\begin{abstract}
In environments such as core-collapse supernovae, neutron star mergers, or the early universe, where the neutrino fluxes can be extremely high, neutrino-neutrino interactions are appreciable and contribute substantially to their flavor evolution. Such a system of interacting neutrinos can be regarded as a quantum many-body system, and prospects for nontrivial quantum correlations, i.e., entanglement, developing in a gas of interacting neutrinos have been investigated previously. In this work, we uncover an intriguing connection between the entropy of entanglement of individual neutrinos with the rest of the ensemble, and the occurrence of spectral splits in the energy spectra of these neutrinos, which develop as a result of collective neutrino oscillations. In particular, for various types of neutrino spectra, we demonstrate that the entanglement entropy is highest for the neutrinos whose locations in the energy spectrum are closest to the spectral split(s). This trend demonstrates that the quantum entanglement is strongest among the neutrinos that are close to these splits, a behavior that seems to persist even as the size of the many-body system is increased.

%We calculate the quantum mechanical many-body   entanglement of neutrinos emerging from collective neutrino oscillations at different energies. To quantify the entanglement we use measures such as the entanglement entropy, the Bloch vector of the reduced density matrix, and logarithmic negativity. Additionally, we present comparisons between the evolution of the neutrino state in the many-body picture vs the mean-field limit, for different initial conditions.
% {We investigate the importance of going beyond the mean-field approximation in the dynamics of collective neutrino oscillations. To expand our understanding of the coherent neutrino oscillation problem, we apply concepts from {many-body physics} and quantum information theory. Specifically, we use measures of nontrivial correlations (otherwise known as ``entanglement'') between the constituent neutrinos of the many-body system, such as the entanglement entropy and the Bloch vector of the reduced density matrix.  The relevance of going beyond the mean field is demonstrated by comparisons between the evolution of the neutrino state in the many-body picture vs the mean-field limit, for different initial conditions. }
%
\end{abstract}

\keywords{Suggested keywords}%Use showkeys class option if keyword
\maketitle

\section{Introduction} \label{section:intro}

Several decades of theoretical and experimental work dedicated to the solar neutrinos culminated with an explanation of the measured distortions of the solar neutrino spectrum in terms of adiabatic and non-adiabatic level crossings~\cite{Wolfenstein78,Mikheyev85,Bethe:1986ej,Haxton:1986uq,Haxton:1987yq,Parke:1986jy}. In the denser media inside supernovae and neutron-star mergers, where neutrinos interact not only with the background particles but also among themselves, more complex phenomena take place.  While solar neutrino oscillations can be primarily described by one-body evolution in a potential governed by coherent forward scattering on background particles, describing neutrino flavor evolution within supernovae and neutron-star mergers requires solving a quantum many-body problem involving nonlinear flavor-dependent forward scattering  among  neutrinos and  inelastic  interactions  with  matter  particles  destroying  coherence~\cite{Fuller87,Notzold:1988fv,Pantaleone:1992eq,Pantaleone92,Sigl:1993fr,Raffelt:1993fj,Raffelt:1993kx,Balantekin:2006tg,Volpe:2013uxl,Vlasenko:2014lr}. Even when one ignores those inelastic interactions, which are subdominant for example sufficiently away from the neutrinosphere in a supernova, one still needs to deal with a Hamiltonian that exhibits an interplay of one- and two-body interaction terms.  There are significant implications of these collective neutrino oscillations in astrophysics: since neutrinos  play  an  essential  role  in  the  supernova  explosions  and  nucleosynthesis~\cite{Fuller:1992eu,Qian:1993dg,Fuller:1993ry,Fuller:1995qy},  neither the  explosion  mechanism  nor  the  nucleosynthetic  output  can  be reliably  predicted  unless  all aspects of the neutrino flavor evolution problem are understood. The same physics also  affects the interpretation of the supernova neutrino  signals in terrestrial detectors. 

In an interacting quantum system with $N$ particles, the size of the Hilbert space typically scales exponentially with $N$. Therefore, in order to study systems with large numbers of particles, various simplifying approaches such as the \lq\lq mean-field\rq\rq\ approximations are frequently adopted. In particular, the mean-field approximations explicitly forbid quantum correlations amongst the constituent particles, thereby reducing the scaling of the effective Hilbert space from exponential to linear. Such approximations have paved the way for extensive numerical treatments of various collective phenomena exhibited by systems of oscillating neutrinos in dense environments (see., e.g., the reviews in Refs.~\cite{Duan:2009cd,Duan:2010fr,Chakraborty:2016a,Tamborra:2020cul}, and the references therein). Whether beyond-the-mean-field effects could have significant implications for the neutrino flavor evolution in these environments remains an interesting and open question. To this end, several exploratory studies have been conducted to investigate the behavior of interacting neutrino systems where inter-particle quantum correlations are permitted~\cite{Friedland:2003dv, Bell:2003mg, Friedland:2003eh, Friedland:2006ke, Balantekin:2006tg, Pehlivan:2011hp, Volpe:2013uxl, Birol:2018qhx, Cervia:2019nzy, Patwardhan:2019zta, Rrapaj:2019pxz, Cervia:2019res, Colombi:2020egf, Roggero:2021asb, Roggero:2021fyo, Hall:2021rbv, Yeter-Aydeniz:2021olz}. Recent interest in this problem has also been spurred by the prospect of simulating such systems using quantum computers~\cite{Hall:2021rbv,Yeter-Aydeniz:2021olz}.

In our previous work we described a procedure for obtaining exact eigenvalues and eigenstates of a many-body neutrino Hamiltonian using a method based on Richardson-Gaudin technique (also known as the Bethe-Ansatz technique), in the two-flavor, single-angle approximation~\cite{Patwardhan:2019zta}. Subsequently, we showed how one could use these eigenvalues and eigenstates to compute the evolution of the many-body neutrino state in the adiabatic limit, using the principle of homotopy continuation, for a variety of initial conditions in flavor~\cite{Cervia:2019res}. We compared the evolution of both the \lq\lq mean-field\rq\rq\ and the many-body density matrices for systems with number of neutrinos $N\le 9$, where the time-dependence of the $\nu$-$\nu$ interaction strength was taken to mimic the bulb model of a core-collapse supernova~\cite{Cervia:2019res}. In the mean-field case, each neutrino interacts separately with the mean field and in the $2 \times 2$ density matrix of each neutrino all the many-particle correlations vanish.  A reliable measure to identify the effect of correlations is the entanglement entropy, which should be zero for the mean-field calculations, but non-zero (albeit bounded) for many-body calculations. 

In the calculations reported in Ref.~\cite{Cervia:2019res} we observed that the entanglement entropy could already reach  to its nearly maximal value, $\log(2)$, with a small number of neutrinos starting with various initial flavor states. Understanding the behavior of the entanglement entropy is crucial to ascertain the validity of the mean-field approximation for the very large number of neutrinos present in core-collapse supernovae and neutron-star mergers. Hence, one of the goals of this paper is to present calculations with an increased number of neutrinos. In doing so, we also identify a very intriguing connection between the entanglement entropy and the \lq\lq spectral splits\rq\rq~\cite{Duan06a,Duan06b,Raffelt07,Raffelt:2007,Duan07a,Duan07b,Duan07c,Fogli07,Duan:2008qy,Dasgupta:2008qy,Dasgupta:2009mg,Dasgupta:2010cd,Friedland:2010yq,Galais:2011gh,Pehlivan:2016lxx,Birol:2018qhx}, which are a commonly occurring phenomenon in systems that exhibit collective neutrino oscillations, even in the mean-field limit. As we illustrate below, the largest values of entanglement entropies occur for neutrinos with energies closest to the spectral split energies.

We introduce the problem and describe the  formalism we use in Section~\ref{sec:formalism}. Section~\ref{sec:evolution} contains a description of the numerical treatment of the many-body evolution. {Our results establishing the connection between entanglement entropy and the location(s) of the spectral split(s) are given in Section~\ref{sec:results}}. We present the analytical inequalities pertaining to the connection between entanglement entropy and the polarization vectors in Section~\ref{sec:traces}. Section~\ref{sec:conclusions} includes brief conclusions.

\section{Formalism}
\label{sec:formalism}

Collective neutrino oscillations in two flavors can be minimally described by the time-dependent Hamiltonian~\cite{Balantekin:2006tg,Pehlivan:2011hp,Birol:2018qhx,Balantekin:2018mpq,Patwardhan:2019zta,Cervia:2019res}
\begin{equation}
    H(t) = -\sum_{\omega}\omega {J}_{\omega}^z + \mu(t) \sum_{\substack{\omega,\omega'\\\omega'\neq\omega}} \vec{J}_\omega\cdot\vec{J}_{\omega'},
    \label{eq:saham}
\end{equation}
in the single-angle approximation{, where neutrinos having the same energy but moving along different trajectories are assumed to have identical flavor evolution. In many cases, this approximation can qualitatively reproduce many of the same behaviors that are observed in the results of more sophisticated multi-angle treatments~\cite{Duan06a,Duan06b,Esteban-Pretel:2007all,Fogli07}\footnote{There are situations where the differences between single-angle and multi-angle calculations can be important (see, e.g., Refs.~\cite{Mirizzi:2010uz,Mirizzi:2013rla,Mirizzi:2013wda,Raffelt:2013rqa,Duan:2014gfa,Abbar:2015mca}, and also the extensive recent literature on fast neutrino oscillations---Ref.~\cite{Tamborra:2020cul} and references therein).}. Here we adopt the single-angle approximation for simplicity, since our focus here is not on trajectory-dependent effects but rather on the role of quantum correlations. {We describe the neutrinos as interacting plane waves (which is quantum mechanically consistent with the choice of assigning them well-defined discrete momenta or oscillation frequencies). Such an approach has been adopted previously in literature~\cite{Bell:2003mg,Friedland:2003eh,Friedland:2006ke}, and has been shown to be adequate for capturing coherent effects~\cite{Friedland:2003eh,Friedland:2006ke}. A calculation that also takes into account incoherent effects would require careful consideration of the fact that individual neutrino trajectories may cross only once per pair, but we regard this to be beyond the scope of this current work.}

In the above equation, $\omega =\delta m^2/(2E)$ is the oscillation frequency in vacuum of a neutrino with energy $E$, where $\delta m^2$ is the difference between the two mass-squared eigenvalues. $\mu = (\sqrt{2}G_F/V) \, D$ parametrizes the $\nu$-$\nu$ interaction strength, where $G_F$ is the Fermi coupling constant, $V$ is the quantization volume and $D$ is a time-dependent geometric factor arising from averaging over the intersection angles of the various neutrino trajectories in the single-angle approximation.
Since the number densities decrease as the neutrino many-body gas expands, and the average over the intersection angles can also be time-dependent, $\mu(t)$ is, in general, explicitly time-dependent. Note that, for simplicity, here we have left out neutrino interactions with background matter, since those can be represented by a one-body interaction term not too dissimilar to the first term in Eq.~\eqref{eq:saham}.

We write the Hamiltonian in Eq.~\eqref{eq:saham} in terms of the neutrino flavor isospin operators in the mass basis:   
\begin{align}
    J_\omega^z &= \frac12 (c_{1\omega}^\dagger c_{1\omega} - c_{2\omega}^\dagger c_{2\omega}),\\
    J_\omega^+ &= c_{1\omega}^\dagger c_{2\omega} = (J_\omega^-)^\dagger,
\end{align}
where $c_{i\omega}^\dagger$ and $c_{i\omega}$ are the creation and annihilation operators of a neutrino mass eigenstate $\ket{\nu_i}$ with label $\omega$. In our calculations we assume that there is exactly one neutrino at each bin in which case the operators above can be represented by Pauli matrices: i.e., $J_\omega^k = \sigma_k/2$. 
The $N$-neutrino many-body state $\ket{\Psi(t)}$  
satisfies the equation 
\begin{equation}
    i\frac{\mathrm{d}}{\mathrm{d}t}\ket{\Psi(t)} = H(t)\ket{\Psi(t)},
    \label{eq:schrod}
\end{equation}
where $H(t)$ is given in Eq.~\eqref{eq:saham}. This state is a pure quantum state in the sense that the associated density matrix $\rho = \ket{\Psi(t)}\!\bra{\Psi(t)}$ satisfies the condition ${\rm Tr} \> \rho^2 =1$. 

One can introduce the polarization vector as 
\begin{equation}
    \vec P_\omega = 2 \> {\rm Tr} \> (\rho\vec J_{\omega}),
    \label{eq:mb-polar1}
\end{equation}
where $\vec{J}_{\omega}$ is the corresponding weak isospin operator for that neutrino. To explore the degree of entanglement between different neutrinos one introduces a reduced density matrix for the neutrino with label $\omega$ by tracing over all other neutrinos with label $\omega'$ not equal to $\omega$: 
\begin{equation}
    \rho^{\rm (red)}_\omega = {\rm Tr}_{\omega' (\neq \omega)} \rho. 
    \label{eq:reduced_density_mat}
\end{equation}
This reduced density matrix is a $2\times 2$ matrix and can be written in terms of the Pauli matrices as 
\begin{equation}
    \rho^{\rm (red)}_\omega  = \frac12 ( \mathbb{I} + \vec{\sigma} \cdot \vec{P}_\omega ),
\end{equation}
where $\mathbb{I}$ is the $2 \times 2$ identity matrix and $\sigma_j$ are the Pauli spin matrices. 
Hence the probability of finding an individual neutrino in the mass eigenstate $\ket{\nu_1}$ is
\begin{equation}
    % \mbox{P}
    P_{\nu_1}(\omega) = \frac12(1+P_{z,\omega}) = [\rho_\omega^{\rm (red)}]_{11},
    \label{eq:polarprob}
\end{equation} 
i.e., the $11$ matrix element of the (reduced) density matrix. 

The entanglement entropy between a neutrino with frequency $\omega$ and the rest of the ensemble takes the form 
\begin{eqnarray}
S(\omega) &=& - {\rm Tr}\> \left[ \rho^{\rm (red)}_\omega \log \rho^{\rm (red)}_\omega \right] \nonumber \\
&=& -\sum_{s=\pm}\lambda_{s,\omega}\log \lambda_{s,\omega},
    \label{eq:ent_form}
\end{eqnarray}
where the eigenvalues of the reduced density matrix $\rho_{\omega}^{\rm(red)}$ are given by 
\begin{align}
    \lambda_{\pm,\omega} &= \frac{1}{2}(1\pm|\vec{P}_\omega|) .
    \label{eq:den_mat_eig}
\end{align}
If the neutrino mode $\omega$ is maximally entangled with its environment (comprised of all the other neutrinos), $|\vec{P}_\omega|=0$, and so entanglement entropy $S(\omega) = \log(2)$. 

%The polarization vector for a neutrino with a given $\omega$ can be written as 
%\begin{equation}
%    \vec P(\omega) = 2 \bra{\Psi} \vec J_{\omega} \ket{\Psi},
%    \label{eq:mb-polar}
%\end{equation}

%The above expression holds in both the exact many-body calculation as well 
In the mean-field limit the wave function factorizes into a direct product of individual neutrino wave functions, i.e., $\ket{\Psi_{\rm MF}} = \bigotimes_\omega \ket{\psi_\omega}$, and the polarization vectors are given by $\vec {\cal P}_\omega = 2 \bra{\psi_\omega} \vec J_{\omega} \ket{\psi_\omega}$, using only the mean-field state $\ket{\psi_\omega}$.  As a consequence, these polarization vectors satisfy the condition $|\vec {\cal P}_\omega|=1$ (implying $S(\omega)=0$ exactly). 
In this sense, the entanglement entropy probes deviations from the mean-field limit due to many-body effects. 
%In the mean-field treatment, the evolution of the $N$-body neutrino system can then be described using a set of $N$ differential equations, each describing the evolution of one neutrino. 
Finally we note that in the mean-field limit the polarization vectors $\vec{{\cal P}}_\omega$ satisfy
%\footnote{Each polarization vector is time-dependent, as can be seen from the definition in Eq.~\eqref{eq:mb-polar1}. This time dependence is suppressed in our notation above for clarity.} 
the evolution equations~\cite{Duan:2010fr}
\begin{equation}
    \frac{\mathrm{d}\vec {\cal P}_\omega}{\mathrm{d}t} = \omega \vec B \times \vec {\cal P}_\omega + \mu(t) \left[ \sum_{\omega'} \vec {\cal P}_{\omega'} \right] \times \vec {\cal P}_\omega,
\end{equation}
where in the mass basis $\vec B = (0,0,-1)$.%, and $\mu(r)$ parametrizes the $\nu$-$\nu$ interaction strength. [and 

%Recall values of entanglement entropy for a single neutrino with respect to the rest of the ensemble are bound between $0$ and $\log(2)$.

\section{Many-body evolution}
\label{sec:evolution}

%Short: We use RK4 to evolve state with sparse matrix rep of Hamiltonian, use sparse matrices to calculate polarization and entanglement entropy.

We consider a system comprised by neutrinos initially in definite flavor states propagating in an isotropic geometry as in the bulb model. We assume that the neutrino flavor field can be represented by a steady-state configuration, wherein the neutrino flavor state at any given location does not explicitly depend on time. As a result, the Schr\"odinger equation can be re-written in terms of the variable of integration $r$ (radius) instead of $t$. The initial many-body state has the form $\ket{\Psi}=\bigotimes_{j=1}^N\ket{\nu_{\alpha_j}}$, where $\alpha_j=e$ or $x$ for each $j$, and evolves according to Eq.~\eqref{eq:schrod} with the time-dependent Hamiltonian in Eq.~\eqref{eq:saham}. The neutrinos are chosen to have discrete, equally spaced vacuum oscillation frequencies $\omega_j = j\omega_0$, for $j=1,\ldots,N$ (where $\omega_0$ is an arbitrary reference frequency), such that each oscillation frequency is occupied by a single neutrino.

For the $\nu$-$\nu$ interaction strength $\mu$, we use the following form that is motivated by the single-angle neutrino bulb model~{\cite{Duan06a,Duan:2010fr}}:
\begin{equation}
	\mu(r) = \frac{G_F}{\sqrt{2} \, V(R_\nu)}\bigg[1-\bigg(1-\frac{R_\nu^2}{r^2}\bigg)^{1/2}\bigg]^2,
	\label{singleMu}
\end{equation}
where $V(R_\nu)$ is the quantization volume for the neutrinos at the neutrinosphere surface $R_\nu$. For definiteness, we choose in our calculations the values $R_\nu = 32.2 \, \omega_0^{-1}$, and $\mu(R_\nu) = 3.62 \times 10^4 \, \omega_0$. Subsequently, our starting radius for the evolution was chosen to be $r_0 = 210.64 \, \omega_0^{-1}$, so as to have $\mu(r_0) = 5\,\omega_0$. {These choices are the same as is Ref.~\cite{Cervia:2019res} and reasonably mimic the physical conditions in a core-collapse supernova environment. 
The quantization volume is related to the neutrino number densities as $n_\nu = N/V$, where $N$ is the total number of neutrinos in the system under consideration\footnote{This would suggest that the choice of initial value of $\mu$ should vary with $N$ in order to have the same initial number density $n_\nu$ in each case. In practice, however, the results do not qualitatively depend on the choice of initial $\mu$, as long as the system is evolving smoothly from a regime where $G_F n_\nu \gtrsim \omega_0$ to one where $G_F n_\nu \ll \omega_0$. As a result, for ease of numerical implementation, we choose to start our computations from the same initial value $\mu = 5$, regardless of $N$.}.
The large initial values of $\mu$ arise as a result of large neutrino densities (small quantization volumes) in these environments.} 

In order to transform between the flavor and mass basis, we have chosen to explore two different examples of vacuum mixing angles: (i) $\theta = 0.161$ (approximately equal to $\theta_{13}$) and (ii) $\theta = 0.584$ (approximately equal to $\theta_{12}$). These choices are made in order to explore the dependence of our results on the mixing angles. {In addition, we also performed calculations for a smaller mixing angle $\theta = 0.01$, which would be a typical value of the matter-suppressed mixing angle at our starting radius $r_0$. The results of these calculation were found to be qualitatively similar to those for $\theta = 0.161$, and therefore we omit them from this paper in the interest of brevity.}

\subsection{Numerical methods}

The state $\ket{\Psi(t)}$ is calculated via the classical fourth order Runge-Kutta (RK4) method, accurate through order $(\delta t)^4$ at each time step $\delta t$:
\begin{align}
    \ket{\Psi(t+\delta t)} &= \ket{\Psi(t)}+\frac{1}{6}\delta t\sum_{i=1}^4\ket{k_i(t)} + O[(\delta t)^5]; 
    \label{eq:RK4sum} \\
    \ket{k_1(t)} &= H(t)\ket{\Psi(t)},\\
    \ket{k_2(t)} &= H\bigg(t+\frac{1}{2}\delta t\bigg)\bigg(\ket{\Psi(t)}+\frac{1}{2}\delta t\ket{k_1(t)}\bigg),\\
    \ket{k_3(t)} &= H\bigg(t+\frac{1}{2}\delta t\bigg)\bigg(\ket{\Psi(t)}+\frac{1}{2}\delta t\ket{k_2(t)}\bigg),\\
    \ket{k_4(t)} &= H(t+\delta t)(\ket{\Psi(t)}+\delta t\ket{k_3(t)}).
\end{align}
Note that the normalization of the state $\ket{\Psi}$ is not exactly preserved when evolved approximately according to Eq.~\eqref{eq:RK4sum}; so, between time steps {one can choose to} explicitly re-normalize the resulting wave function, {should step sizes be too large to preserve approximate normalization}.

As we ramp our calculations up to larger and larger values of number of neutrinos, $N$, we must be wary of the scaling of the difference between the extremal eigenvalues in our Hamiltonian, which increases the frequency of the oscillatory nature in the integral of our time evolution operator. To this end, we take time steps of size $0.1\left[\mu \frac{N}{2}(\frac{N}{2}+1)+\sum_\omega|\omega|\right]^{-1}$, where $\mu$ is evaluated at the radius prior to taking this time step. The reason for this choice is that the maximal energy eigenvalue for our system for $\mu\geq0$ is given by $E_{-N/2}\equiv\mu \frac{N}{2}(\frac{N}{2}+1)+\frac{1}{2}\sum_\omega|\omega|$, while $-\frac{1}{2}\sum_\omega|\omega|$ serves as a lower bound for the lowest energy eigenvalue (and is equal to the lowest energy eigenvalue for $\mu = 0$). Our chosen step-size is dynamically adjusted at each time step so that it remains inversely proportional to the difference between these two bounds as $\mu$ changes.

By comparing with the Bethe-ansatz/homotopy-continuation based method presented in Ref.~\cite{Cervia:2019res} for evolving a many-body state with the Hamiltonian $H(t)$, we verified for $N\leq9$ that using Runge-Kutta methods to approximate the evolved state to order $(\delta t)^5$ with the appropriately chosen $\delta t$ produces accurate results for the wave function even after evolving over many time steps. When compared with results obtained using that method, the value of each coefficient in the wave function, $\braket{j|\Psi(t)}$ ($j=0,\ldots,2^N-1$) was found to be discrepant at a level of $\lesssim10^{-6}$. 

%\emph{Something about Bethe ansatz's numerical instability here:}
Our reason for not persisting with the Bethe-ansatz method for the purposes of the current calculations was that we found, at least in our implementation, it can be numerically unstable for $N \geq 10$. The principle used by that method was to construct the solutions of the Bethe-ansatz equations for an arbitrary value of $\mu$ by smoothly increasing $\mu$ from zero (where the solutions have an easy analytic form), since one could show that the solutions for any $\mu$ are continuously connected to the corresponding solutions for $\mu = 0$. However, for $N \geq 10$, this process did not prove robust, as the solutions demonstrated a tendency to jump amongst one another as the parameter $\mu$ was increased. A more careful treatment of 
solutions to the Bethe-ansatz equations at large $\mu$ will be deferred to a future publication.

This use of RK4 in a sparse-matrix representation permits calculations of the evolved many-body wave function according to a time-dependent Hamiltonian for up to $N=16$ on a personal computer. In implementing the sparse representation in our own calculations, submodules from the SPARSKIT Fortran 90 library~\cite{Saad94sparskit:a} for performing operations with sparse matrices are utilized. However, the ability to store the entire many-body Hamiltonian eventually becomes inhibited by limitations on memory as $N$ increases, as its matrix in the mass basis contains ${\cal O}(N^2 \, 2^N)$ nonzero elements. Additionally, the computation time of this evolution similarly scales exponentially in $N$ according to this procedure. % {\color{red} Comment on exploration of tensor network methods in a future publication?}

\section{Results} \label{sec:results}

Using the methods outlined in the previous section, we performed numerical integration of interacting neutrino ensembles, for total neutrino numbers ranging from $N = 2$ to $N=16$, and for various different initial conditions in flavor. To begin with, we wanted to investigate how the amount of entanglement scales with $N$---for this purpose, we picked a test case wherein the initial state consists of all electron-flavor neutrinos, i.e., $\ket{\Psi(r_0)} = \bigotimes_{j=1}^N \ket{\nu_{e,\omega_j}}$. The results of our calculations with this particular initial condition are shown in Figs.~\ref{fig:PzWN}--\ref{fig:Polar16Spec001}, and they essentially amount to an extension of Fig.~1a from our previous paper (Ref.~\cite{Cervia:2019res}).

Subsequently, we explored the evolution of neutrino ensembles with different initial conditions, where some neutrinos start as $\ket{\nu_e}$ and others as $\ket{\nu_x}$. In Fig.~\ref{fig:lomixspec8}, we reproduce a result previously shown in Ref.~\cite{Cervia:2019res} (albeit using a different numerical method), whereas through the remaining plots shown in Figs.~\ref{fig:Polar8Spec016}--\ref{fig:Polar16Spec016}, we extend the results of Ref.~\cite{Cervia:2019res} in various ways---e.g., by changing to a large mixing angle (Fig.~\ref{fig:himixspec8}) or by adding more particles (Fig.~\ref{fig:Polar16Spec016}). Through these calculations, we glean a number of interesting insights which are described throughout the remainder of this section.

%{\color{red} Results section probably needs an overview paragraph before we dive into a description of figures.}

\begin{figure*}[htb]
\begin{center}
    \subfloat[\label{fig:lomixN}]{
	    \includegraphics[width=0.48\textwidth]{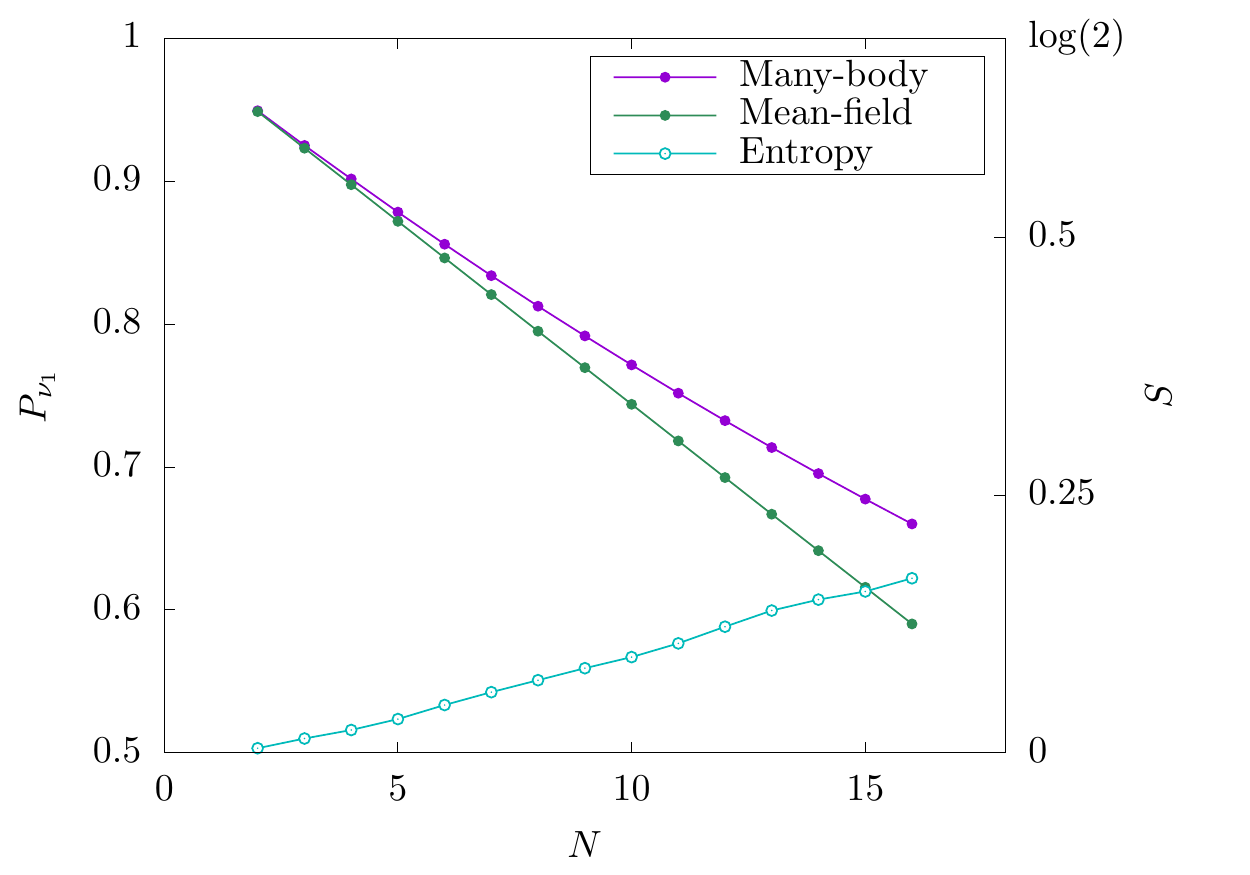}}
	~
	\subfloat[\label{fig:himixN}]{
	    \includegraphics[width=0.48\textwidth]{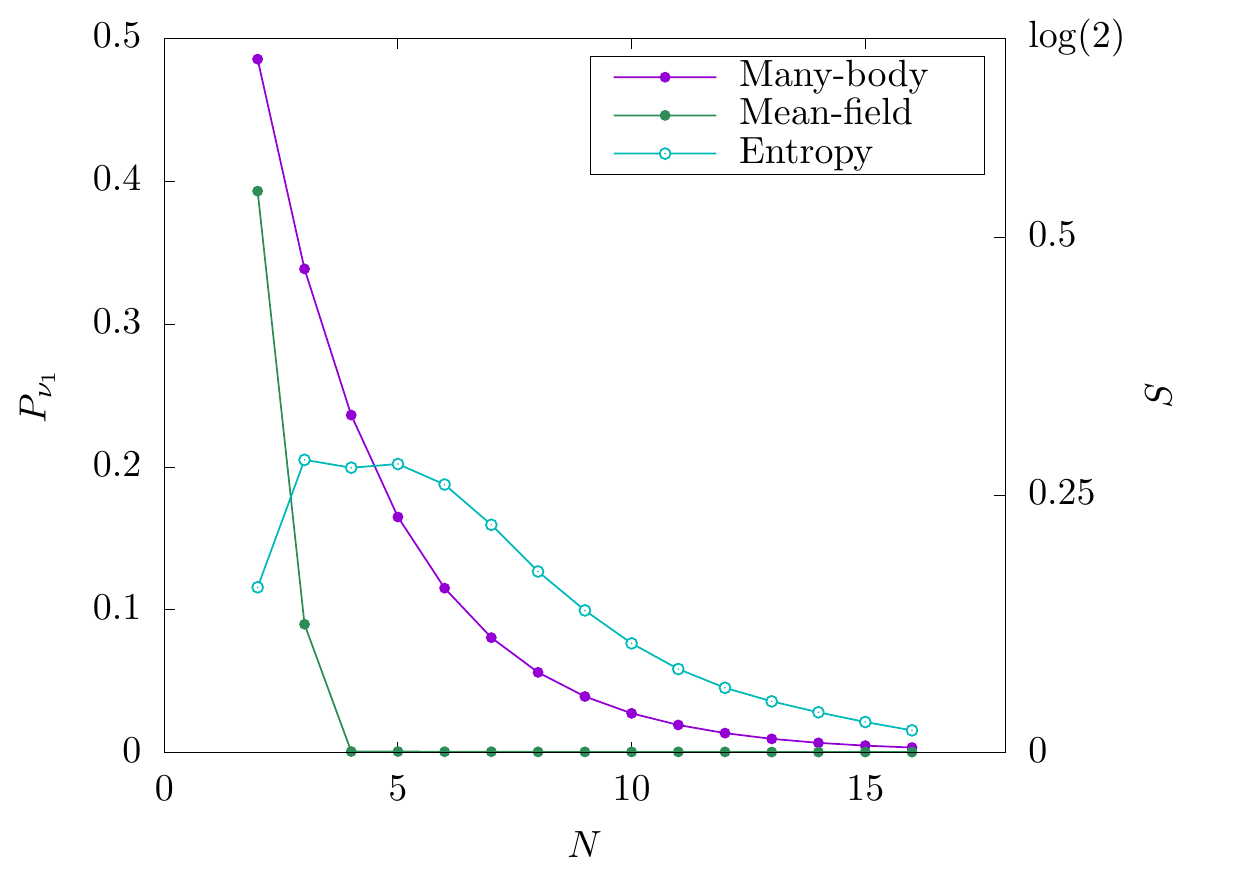}}
\end{center}
    \caption{A comparison between many-body (purple) and mean-field (green) calculations of the flavor evolution of systems with a total neutrino number ranging from, $N=2$ to $N=16$, each starting from an initial configuration $\ket{\nu_e, \ldots, \nu_e}$ at $\mu_0 = 5\omega_0$. Shown here are the asymptotic (i.e., at $r\gg R_\nu$, or equivalently, $\mu \ll \omega_0$) values of $P_{\nu_1}(\omega_N)$, i.e., the probability of detecting the neutrino  with the highest oscillation frequency $\omega_N$ in a $\ket{\nu_1}$ mass eigenstate at a large distance. Also shown in each plot is the asymptotic value of the entropy of entanglement (cyan) between the neutrino with frequency $\omega_N$ and the remaining neutrinos in the ensemble. Figures~\ref{fig:lomixN} and \ref{fig:himixN} portray the results of calculations with mixing angles $\theta=0.161$ and $\theta=0.584$, respectively.}
    \label{fig:PzWN}
\end{figure*}

Figure~\ref{fig:PzWN} shows the results of computing $P_{\nu_1}(\omega_N)$, i.e., the probability of the neutrino in bin $\omega_N$ ($= N\omega_0$, the highest vacuum oscillation frequency) being found in the $\ket{\nu_1}$ mass eigenstate, at a final time where $\mu\ll\omega_0$, for systems with neutrino numbers ranging from $N=2$ to $N=16$. As mentioned previously, we performed two sets of calculations, with mixing angles $\theta = 0.161$ and $\theta=0.584$, represented in Figs.~\ref{fig:lomixN} and \ref{fig:himixN}, respectively. At each $\omega$, this probability $P_{\nu_1}(\omega)$ is related to the $z$-component of the mass-basis polarization vector, $P_{z,\omega}$, by Eq.~\eqref{eq:polarprob}. 

{For the small mixing angle, we find that, among all the neutrinos, the ones with frequencies $\omega_{N-1}$ and $\omega_N$ exhibit the highest amount of entanglement with the rest of the ensemble, as quantified by their respective entanglement entropies. Correspondingly, at these two frequencies, the value of $P_{\nu_1}$ deviates the most from its mean-field predicted value (see, e.g., Fig.~\ref{fig:lomixspec16}). One must note that, for this particular initial condition and mixing angle, the location of the spectral split frequency $\omega_s$ lies in-between $\omega_{N-1}$ and $\omega_{N}$ for the range of values of $N$ considered here. For a system of $N$ neutrinos with an evenly spaced spectrum of oscillation frequencies (as described in Sec.~\ref{sec:evolution}), all initially in the $\nu_e$ flavor state, and with a mixing angle $\theta$, the spectral split will center around the split frequency $\omega_s$ given by\footnote{For a proof, see Refs.~\cite{Birol:2018qhx,Balantekin:2018mpq}, for example.}
\begin{equation}
    \omega_s = \omega_0 \, N \, \cos^2 \theta,
    \label{eq:allesplit}
\end{equation}
as a consequence of the conservation of $J^z=\sum_\omega J_\omega^z$. Since the neutrinos showing the highest degree of entanglement in this case are the ones closest to $\omega_s$, this suggests that it may be instructive to more closely examine the neutrinos close to the split frequencies in various other cases as well. }

In order to test a scenario where the location of $\omega_s$ is further inside the spectrum rather than close to its edge, we also performed a set of calculations with a larger mixing angle, $\theta = 0.584$. For a large mixing angle, we find that the final value of $P_{\nu_1}(\omega_N) \to 0$ as $N$ grows, so that the many-body result converges towards the mean-field predicted value. Correspondingly, the entanglement entropy of this neutrino decreases with growing $N$ as well. Indeed, this behavior is correlated with the location of the spectral split frequency $\omega_s$ moving further inside and away from $\omega_N$ as $N$ is increased. With these trends in mind, it is then natural to consider the behavior of the neutrino modes nearer to the spectral split frequency $\omega_s$ instead. These results are shown in Fig.~\ref{fig:PzWNs}, for small and large mixing angles (Figs.~\ref{fig:lomixs} and \ref{fig:himixs}, respectively).

In general, $\omega_s$ lies somewhere in between two consecutive oscillation frequencies in our discrete spectral grid. Let $N_s:=\omega_s/\omega_0$ be the effective index of a neutrino in the frequency spectrum at which the spectral split is centered. To estimate the values of the relevant physical quantities (such as $P_{\nu_1}$ or $S$) \lq\lq at\rq\rq\ the spectral split location, one must interpolate between the discrete steps in oscillation frequencies in our calculations. For a distribution of neutrinos evenly spaced in oscillation frequencies, we can define an estimated value at the split frequency by linearly interpolating a function of the discrete oscillation frequency spectrum as
\begin{align}
    F(\omega_s) =& (1-(N_s-\lfloor N_s \rfloor))F(\lfloor N_s \rfloor \omega_0) \nonumber \\
    +& (1-(\lceil N_s \rceil-N_s))F(\lceil N_s \rceil \omega_0),
    \label{eq:interpolate}
\end{align}
where $F$ may be $P_{\nu_1}$ or $S$ for example, and where \lq\lq $\lfloor \cdot \rfloor$\rq\rq\ and \lq\lq $\lceil \cdot \rceil$\rq\rq\ respectively represent the floor and ceiling functions. 
%We could give these floor and ceiling frequencies names $\omega_{s1}=\lfloor N_s \rfloor\omega_0$ and $\omega_{s2}=\lceil N_s \rceil\omega_0$?

\begin{figure*}[htb]
\begin{center}
    \subfloat[\label{fig:lomixs}]{
	    \includegraphics[width=0.48\textwidth]{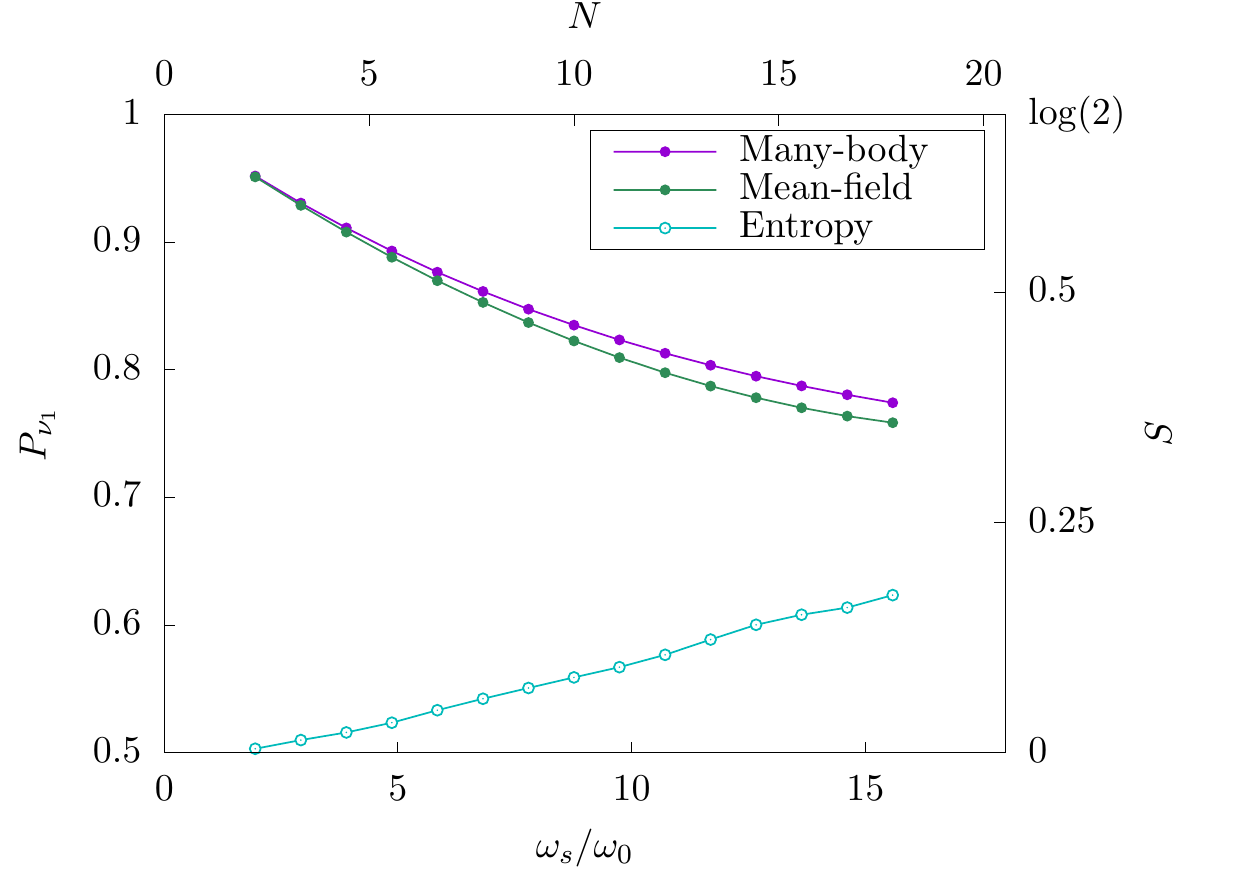}}
	~
	\subfloat[\label{fig:himixs}]{
	    \includegraphics[width=0.48\textwidth]{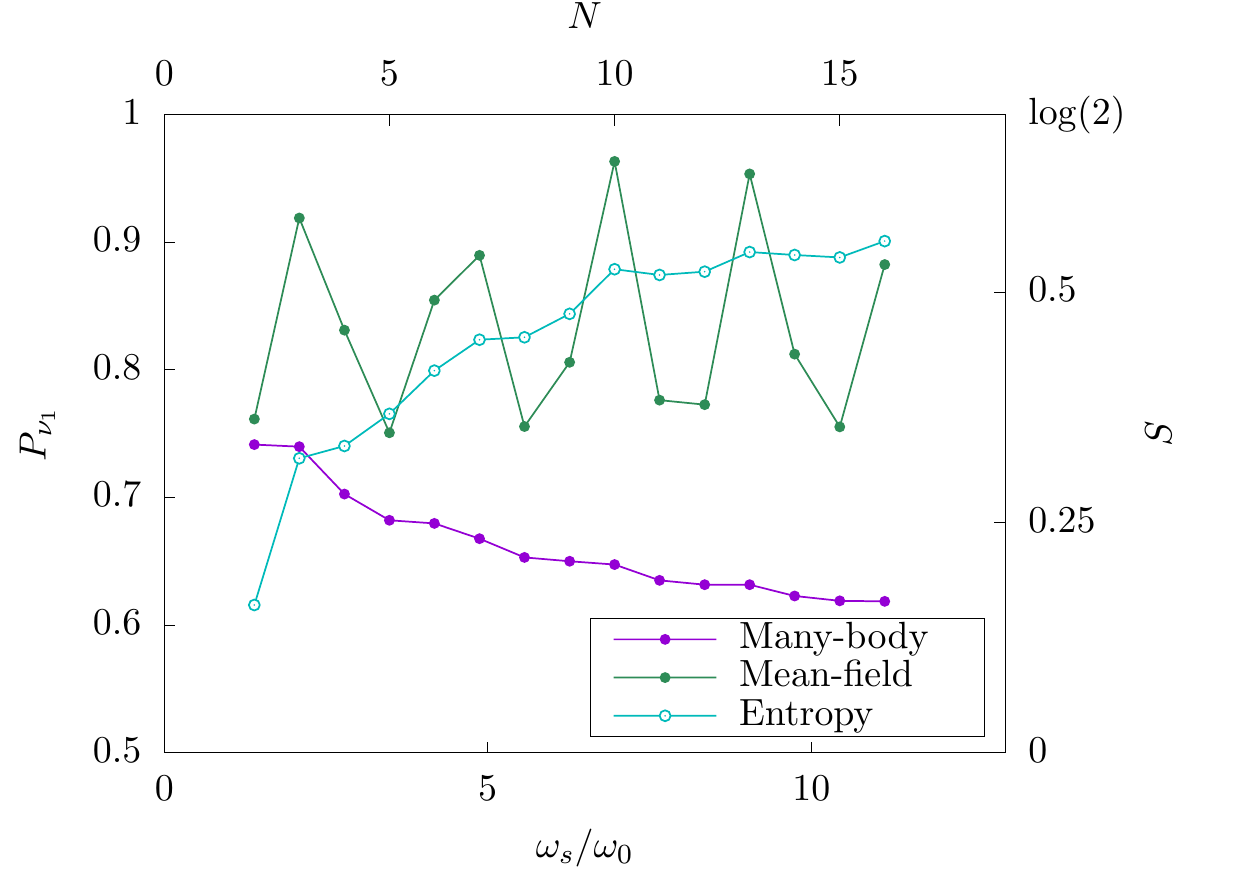}}
    
\end{center}
    \caption{
    %A comparison between many-body and mean-field calculations of the flavor evolution of systems with various numbers of neutrinos, $N$, each starting from an initial configuration $\ket{\nu_e, \ldots, \nu_e}$ at $\mu_0 = 5\omega_0$. 
    Same as Fig.~\ref{fig:PzWN}, but where the asymptotic probabilities and entanglement entropies at evaluated at $\omega = \omega_s$ instead of $\omega_N$, using an interpolation scheme. $P_{\nu_1}(\omega_s)$ is defined as a linear interpolation of the $\ket{\nu_1}$ eigenstate detection probabilities, $P_{\nu_1}(\lfloor N_s \rfloor\omega_0)$ and $P_{\nu_1}(\lceil N_s \rceil\omega_0)$, for the neutrinos with oscillation frequencies on either side of the spectral swap frequency $\omega_s=N_s\omega_0$. The specific linear interpolation scheme used here is described in Eq.~\eqref{eq:interpolate}. Likewise, in each figure, the asymptotic value of the entropy of entanglement at the split frequency, $S(\omega_s)$, is defined as a linear interpolation between the frequencies $\lfloor N_s \rfloor\omega_0$ and $\lceil N_s \rceil\omega_0$ as per Eq.~\eqref{eq:interpolate}. Figures~\ref{fig:lomixs} and \ref{fig:himixs} portray the results of calculation with mixing angles $\theta=0.161$ and $\theta=0.584$, respectively.} 

    \label{fig:PzWNs}
\end{figure*}

\begin{figure*}[htb]
\begin{center}
    \subfloat[\label{fig:lomixspec16}]{
	    \includegraphics[width=0.48\textwidth]{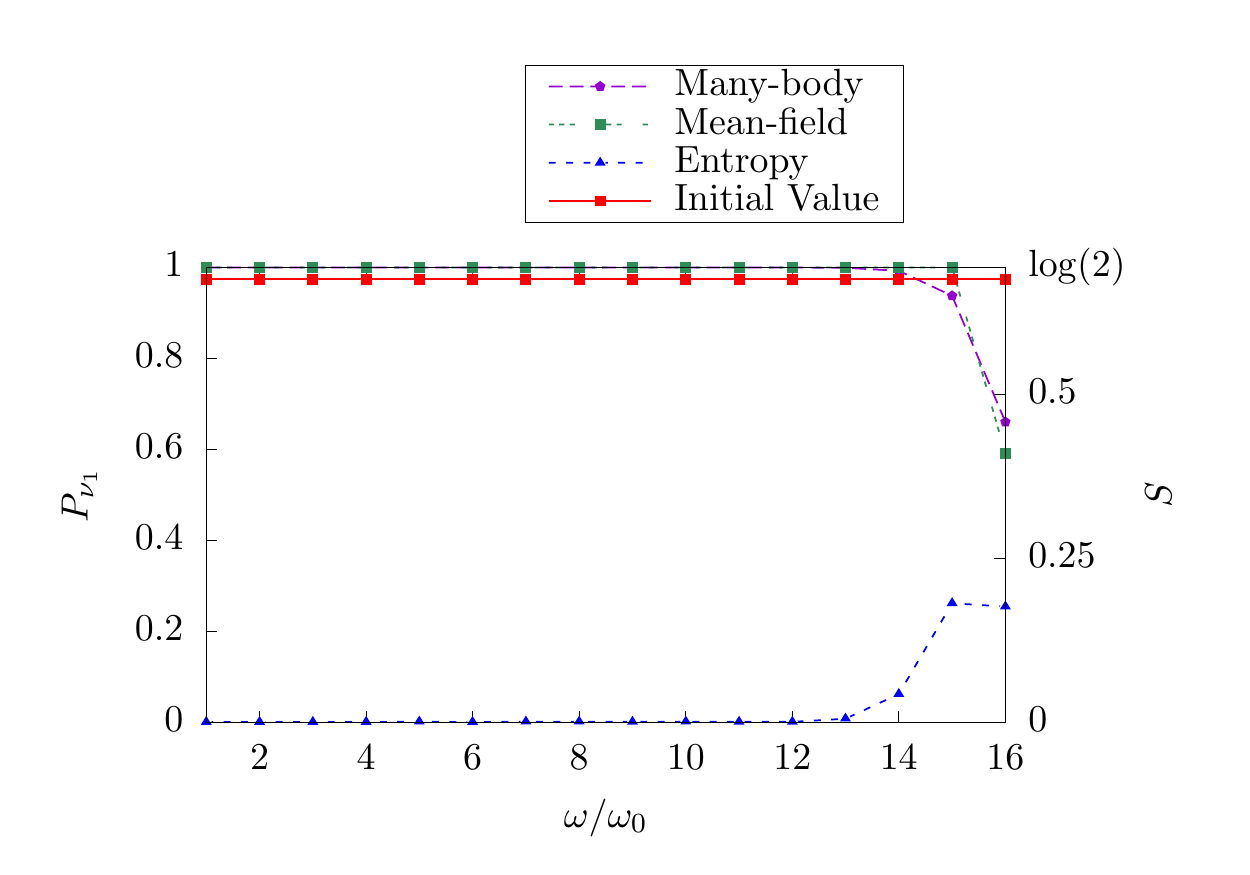}}
	~
	\subfloat[\label{fig:himixspec16}]{
	    \includegraphics[width=0.48\textwidth]{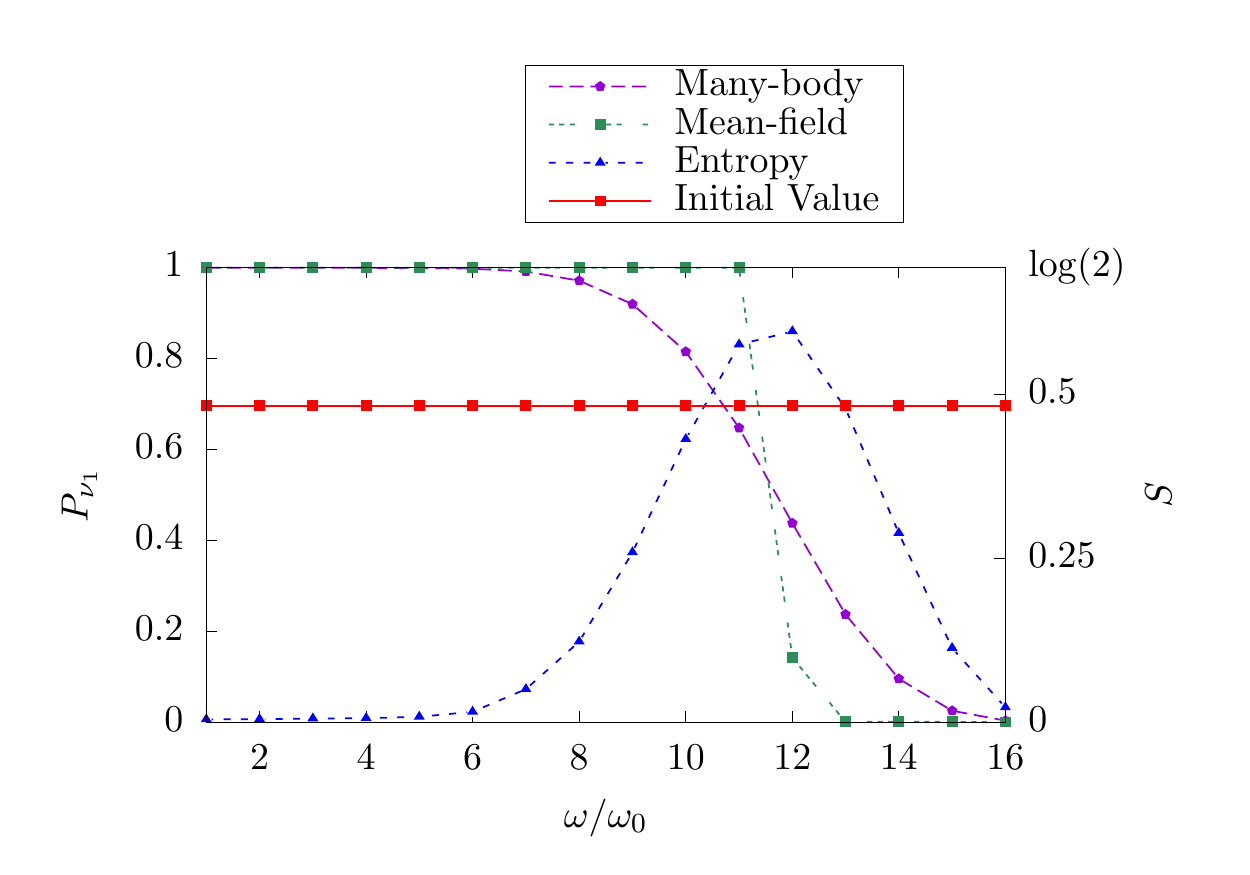}}
    
\end{center}
    \caption{Comparison of the asymptotic (i.e., at $r\gg R_\nu$, or equivalently, $\mu \ll \omega_0$) spectra of $P_{\nu_1}$ vs.~vacuum oscillation frequency $\omega$, i.e., the probabilities of each of the neutrinos being found in the $\ket{\nu_1}$ mass eigenstate. Shown here are the many-body (purple) and mean-field (green) calculations, for a $N=16$ neutrino system with an initial configuration consisting of a $\ket{\nu_e}$ at each of the frequencies $\omega_1,\ldots,\omega_{16}$; the initial values of these probabilities (red) at each frequency are also shown for comparison. Also shown are the entropies of entanglement (blue) between each neutrino and the remaining neutrinos in the ensemble. Figures~\ref{fig:lomixspec16} and~\ref{fig:himixspec16} portray the results of calculation with mixing angles $\theta=0.161$ and $\theta=0.584$, respectively. It can be seen, particularly in Fig.~\ref{fig:himixspec16}, that the entropy of entanglement peaks around the spectral split frequency ($\omega/\omega_0 \approx 11\mbox{--}12$), implying that the neutrinos closest to the split have the strongest entanglement.} 
        % Left: Comparison between mean-field and many-body calculations of the $z$ component of the polarization vector, $P_{z}(\omega_N)$, for the neutrino with highest frequency $\omega_N$. Shown here are the asymptotic (i.e., at $r\gg R_\nu$) values of $P_{z}(\omega_N)$, for systems with various numbers of neutrinos, $N$, each starting from an initial configuration $\ket{\nu_e, \ldots, \nu_e}$ at $\mu_0 = 5\omega_0$ [or, equivalently, at the radius $r_0\approx 6.6 R_\nu$, as per Table~\ref{parameters} and Eq.~\eqref{singleMu}]. Notably, for this configuration, the discrepancy between the mean-field and many-body results grows with increasing $N$, at least for the small values of $N$ examined here. Right: The entanglement entropy $S(\omega_N)$ of the highest frequency neutrino with the rest of the ensemble [calculated using Eqs.~\eqref{evolvedState}--\eqref{densityRed}], as a function of radius $r$, for the same set of systems as in the left panel. 
    %In mean-field theory this entropy is zero. %As with the magnitude of the difference between many-body and mean-field values of $P_z(\omega_N)$ in the left panel,
    %Here the asymptotic values of $S(\omega_N)$ also %appear to 
    %grow with $N$, demonstrating a possible correlation with the discrepancy between many-body and mean-field results.
    %Note that $\mu$ decreases with $r$---here we use the relationship from Eq.~\eqref{singleMu}, which is borrowed from the single-angle bulb model~\cite{Duan:2010fr}.

    \label{fig:Polar16Spec001}
\end{figure*}

\begin{figure*}[htb]
\begin{center}
    \subfloat[\label{fig:lomixspec8}]{
	    \includegraphics[width=0.48\textwidth]{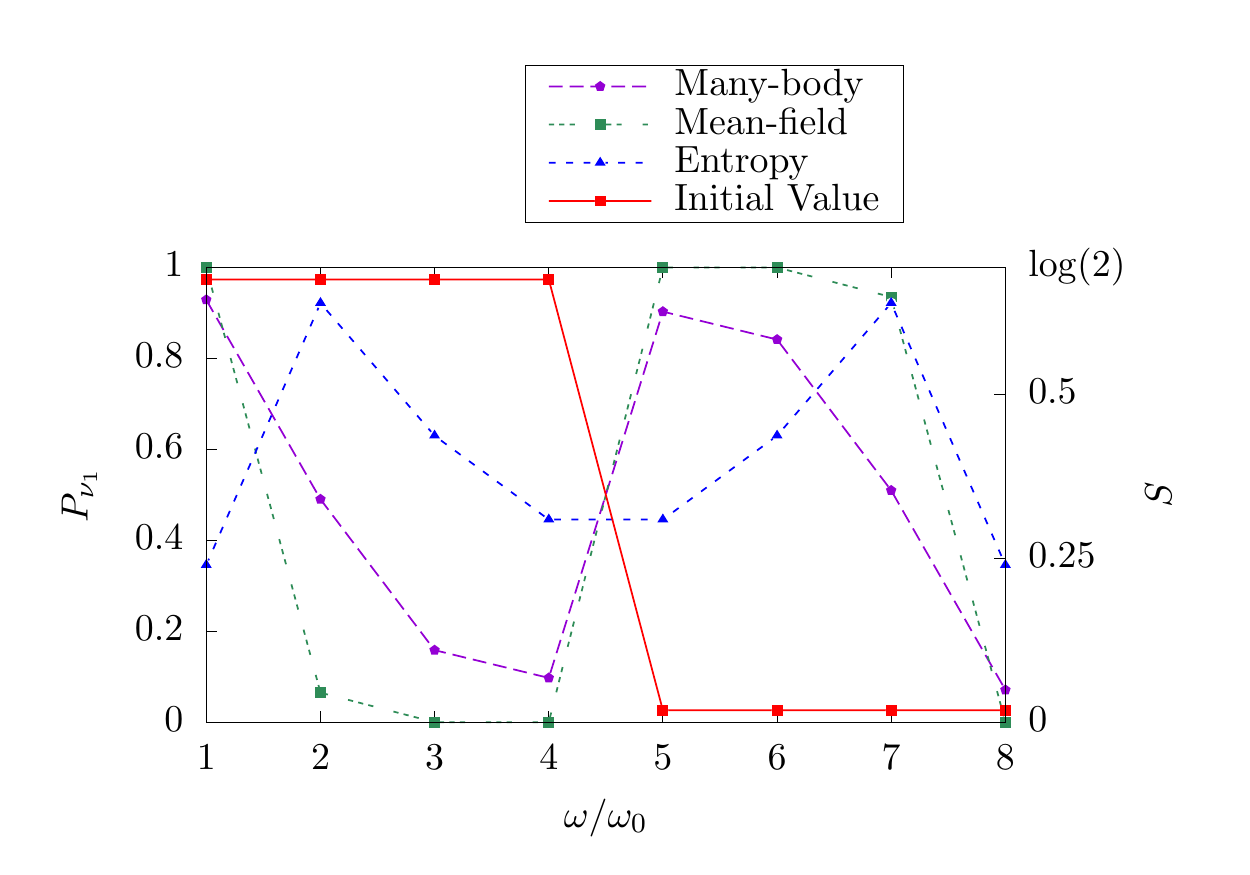}}
	~
	\subfloat[\label{fig:himixspec8}]{
	    \includegraphics[width=0.48\textwidth]{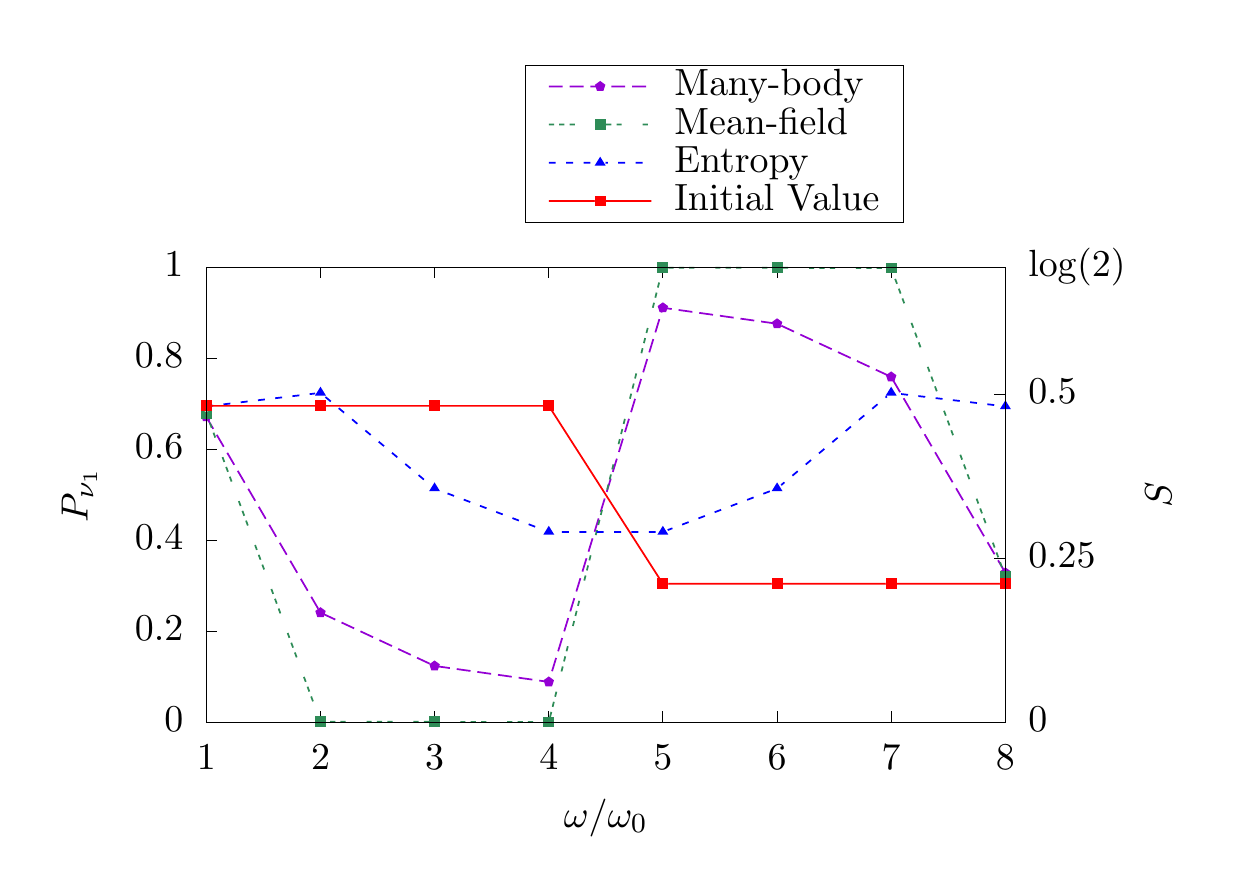}}
    
\end{center}
    \caption{Same as Fig.~\ref{fig:Polar16Spec001}, but for an $N=8$ neutrino system with an initial configuration consisting of a $\ket{\nu_e}$ at each of the frequencies $\omega_1,\ldots,\omega_4$, and a $\ket{\nu_x}$ at each of $\omega_5,\ldots,\omega_8$. Figures~\ref{fig:lomixspec8} and ~\ref{fig:himixspec8} portray the results of calculation with mixing angles $\theta=0.161$ and $\theta=0.584$, respectively. The entropy of entanglement can be seen to peak near the two spectral split frequencies, $\omega/\omega_0 \approx 2$ and $7$. %It can be seen, particularly in Fig.~\ref{fig:himixspec8}, that the entropy of entanglement peaks around the spectral split frequency, implying that the neutrinos closest to the split have the strongest entanglement.
    } 
    \label{fig:Polar8Spec016}
\end{figure*}

\begin{figure*}[htb]
\begin{center}
    \subfloat[\label{fig:lomixspec16_016}]{
	    \includegraphics[width=0.48\textwidth]{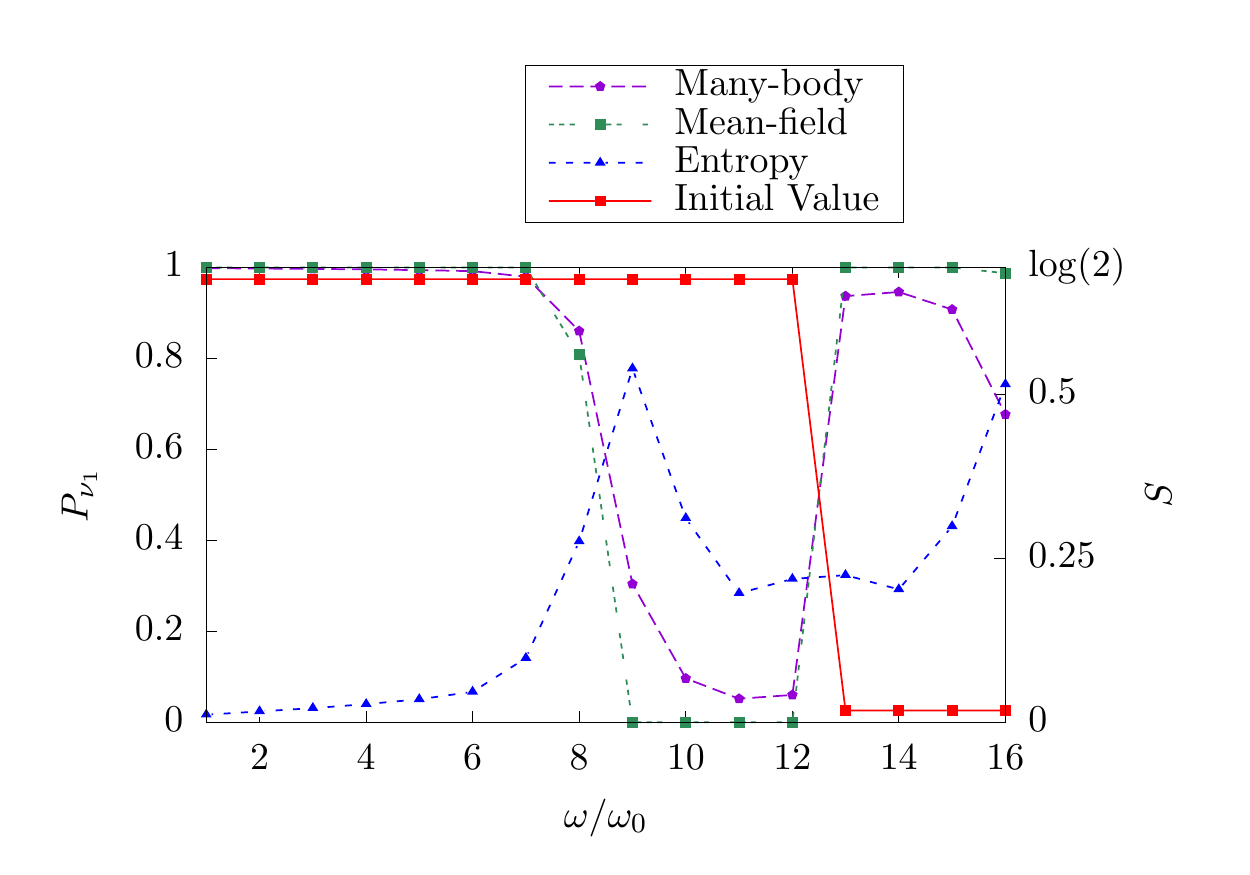}}
	~
	\subfloat[\label{fig:himixspec16_016}]{
	    \includegraphics[width=0.48\textwidth]{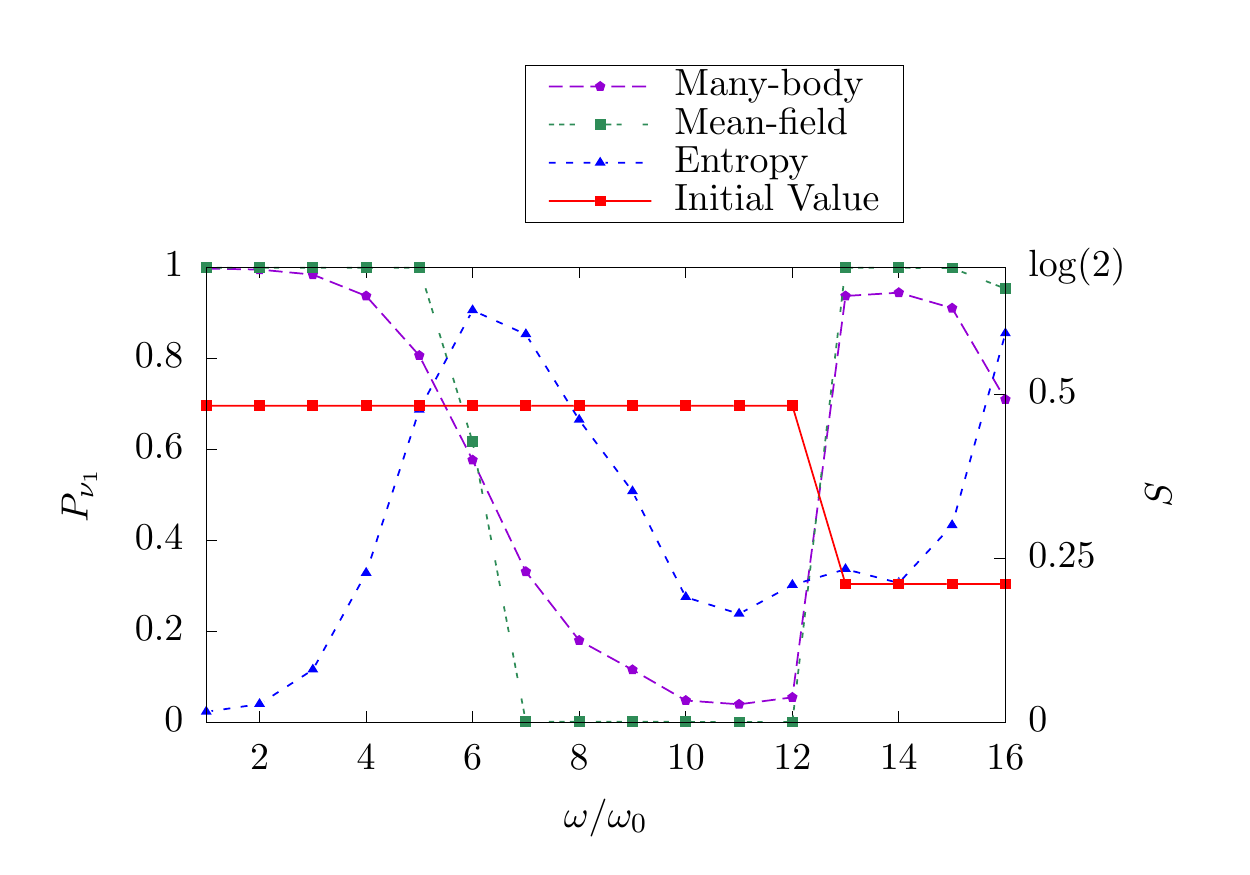}}
	\\
	\subfloat[\label{fig:lomixspec16_256}]{
	    \includegraphics[width=0.48\textwidth]{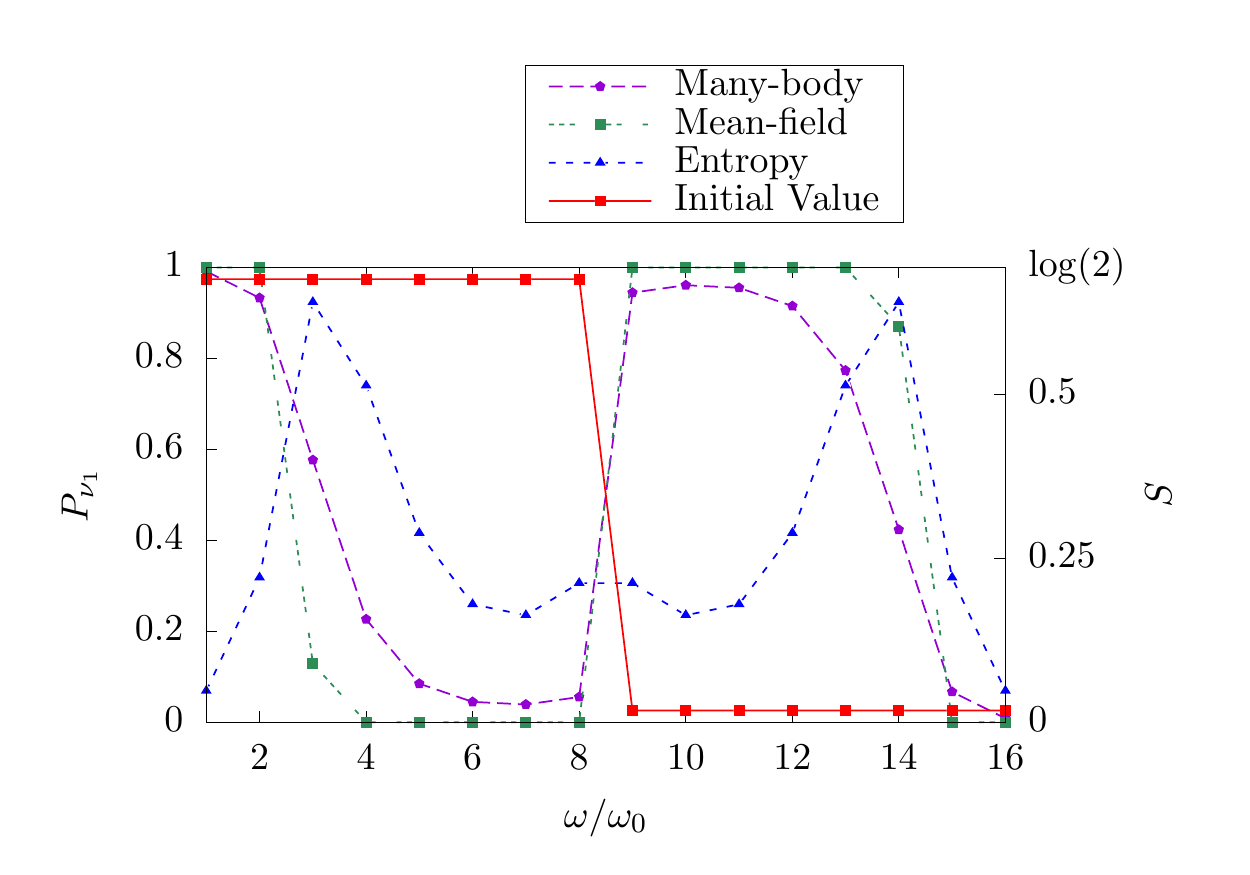}}
	~
	\subfloat[\label{fig:himixspec16_256}]{
	    \includegraphics[width=0.48\textwidth]{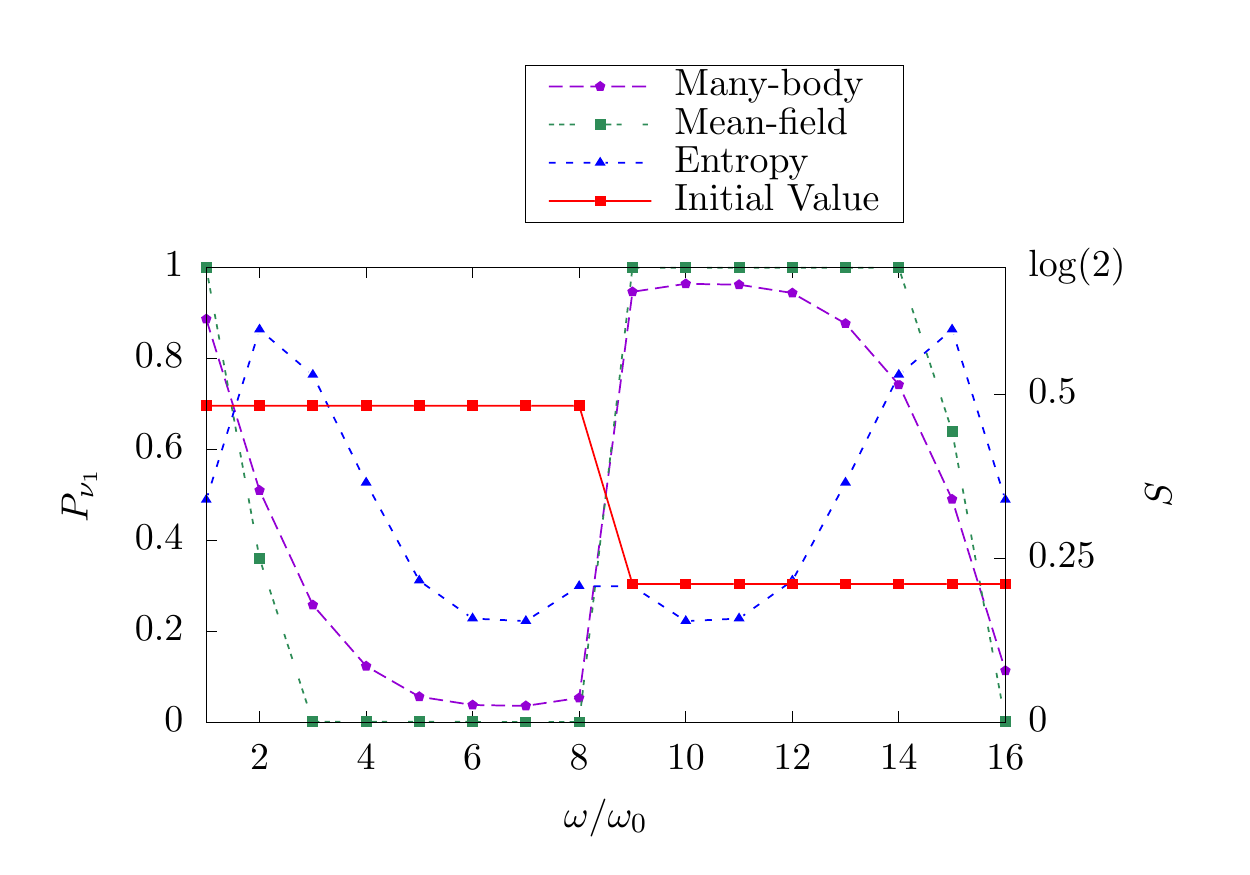}}
\end{center}
    \caption{Same as Fig.~\ref{fig:Polar16Spec001}, but for $N=16$ neutrino systems with different initial configurations: shown in Figs.~\ref{fig:lomixspec16_016} and~\ref{fig:himixspec16_016} are the results for a system with an initial configuration consisting of a $\ket{\nu_e}$ at each of the frequencies $\omega_1,\ldots,\omega_{12}$, and a $\ket{\nu_x}$ at each of $\omega_{13},\ldots,\omega_{16}$. Figures~\ref{fig:lomixspec16_016} and ~\ref{fig:himixspec16_016} portray the results of calculation with mixing angles $\theta=0.161$ and $\theta=0.584$, respectively. Figures~\ref{fig:lomixspec16_256} and~\ref{fig:himixspec16_256} show the corresponding results for $\theta=0.161$ and $\theta=0.584$, respectively, for a system with an initial configuration consisting of a $\ket{\nu_e}$ at each of the frequencies $\omega_1,\ldots,\omega_{8}$, and a $\ket{\nu_x}$ at each of $\omega_{9},\ldots,\omega_{16}$. In each case, the entropy of entanglement appears to peak near the respective spectral split frequencies, e.g., at $\omega/\omega_0 \approx 3$ and $14$ in Fig.~\ref{fig:lomixspec16_256}.} 
    \label{fig:Polar16Spec016}
\end{figure*}

With this convention for defining physical quantities at $\omega_s$, we display the analogous results for the asymptotic ($\mu \ll \omega_0$) evolved values of $P_{\nu_1}(\omega)$ and $S(\omega)$ at $\omega = \omega_s$ in Fig.~\ref{fig:PzWNs}. Here, we find for a small mixing angle that the deviation of final $P_{\nu_1}(\omega_s)$ from the mean-field predicted value grows at a smaller rate in $N$; however, we again find a steady growth in entanglement entropy with $N$. These results are unsurprising, because $\omega_s$ is close to $\omega_N$ for small $\theta$, as per Eq.~\eqref{eq:allesplit}. In contrast, for a large mixing angle, we find that $\omega_s$ is much further from $\omega_N$, and therefore in Fig.~\ref{fig:himixs} we observe an entirely different trend with increasing $N$ for the corresponding $P_{\nu_1}$ and $S$ values, compared to Fig.~\ref{fig:himixN}. The $P_{\nu_1}(\omega_s)$ values for the many-body calculations decrease monotonically with $N$, whereas in the mean-field calculations, one observes sharp oscillatory features in the $P_{\nu_1}(\omega_s)$ vs.~$N$ trend, arising mainly because of the mean-field spectral splits being much sharper, resulting in the interpolation also being less smooth (e.g., see Fig.~\ref{fig:himixspec16}---the $P_{\nu_1}$ values on either side of the split are sharply pulled towards $1$ and $0$ in the mean-field calculations, in contrast with the many-body spectral split, where the trend towards $P_{\nu_1} = 1$ or $0$ on either side of the split is much more gradual). But the key takeaway is that $S(\omega_s)$ continues to grow with $N$, which seems to suggest that increasing the particle number does not result in a decrease of entanglement around the spectral split region.

Figures~\ref{fig:Polar16Spec001}, \ref{fig:Polar8Spec016}, and~\ref{fig:Polar16Spec016} show the results of the asymptotic spectra of $P_{\nu_1}(\omega)$ vs.~$\omega$, i.e., the probabilities for each of the neutrinos in the ensemble to be detected in the $\ket{\nu_1}$ state at far distances where $\mu \ll \omega_0$. Different figures represent neutrino ensembles that started from different initial conditions in flavor. For instance, Fig.~\ref{fig:Polar16Spec001} represents a neutrino ensemble wherein all neutrinos started in electron flavor, whereas Figs.~\ref{fig:Polar8Spec016} and \ref{fig:Polar16Spec016} represent neutrino ensembles where some neutrinos started as $\ket{\nu_e}$ and others as $\ket{\nu_x}$. In each figure, we show the results for two sets of calculations, namely with mixing angles in vacuum of $\theta = 0.161$ and $\theta = 0.584$, respectively. The final state configurations of the neutrino ensembles in all of these calculations exhibit the presence of spectral splits, in the many-body as well as in the mean-field calculations. Unsurprisingly, in each figure, the locations of the spectral splits in the many-body and mean-field calculations coincide with one another, since in either case, the physics behind the origin of these splits is based on the total $J^z$ being a conserved charge of the neutrino Hamiltonian. Generically, across all the results, one can make the following observations:
\begin{enumerate}
    \item The entanglement entropy is maximum for the neutrinos that have frequencies nearest to the spectral split frequencies. This is more robust of a finding than the correlation between entanglement entropy and the deviation in asymptotic values of $P_z$ that was noted in Ref.~\cite{Cervia:2019res}.
    \item The spectral splits generically appear to be much broader in the many-body calculations than in the mean-field calculations. This observation was already noted in Ref.~\cite{Cervia:2019res} for neutrino ensembles with $N \leq 8$, and here the same behavior is manifested even in systems with neutrino numbers up to $N = 16$.
    \item The width of the spectral splits in the many-body calculations seem to depend on the mixing angle $\theta$, unlike in the mean-field case where the splits are always sharp.
    \item {A comparison of Fig.~\ref{fig:lomixspec8} and Fig.~\ref{fig:lomixspec16_256} shows that, as $N$ is increased, the width of the $S(\omega)$ vs.~$\omega$ bell curves does not appear to grow in proportion with the total width of the neutrino spectrum. In fact, the width seems to remain more or less constant, even as $N$ is increased from 8 to 16, suggesting that the entanglement remains localized in the immediate neighborhood of the splits. In other words, in relation to the total width of the neutrino spectrum ($\omega_N - \omega_1$), the width of the splits appears to shrink with increasing $N$. We have verified that this latter assertion holds true even if ($\omega_N - \omega_1$) is held fixed through a rescaling of the oscillation frequencies as $N$ is increased.}
\end{enumerate}
In particular, given the correlation between entanglement entropy and the deviation relative to mean-field calculations, observation 4 suggests that, even as the system gets larger, only the neutrinos that are closest to the split frequency deviate strongly from the mean-field calculations. Such behavior suggests that, in order to scale such computations to systems with large numbers of neutrinos, a hybrid computational approach may be feasible, wherein neutrinos away from the spectral split region(s) are evolved using a mean-field treatment, whereas those closer to the split are treated as true many-body system.

% We would also like to point out that we can calculate with very small mixing angles resulting from suppression by matter effects in a matter-dense environment, i.e.,$\theta=0.01$. (We take an effective mixing angle without time dependence to estimate the behavior of collective oscillations with suppression as large as it is at the radius of $\mu_0$.) In such a calculation, one would find ... 

\section{Trace inequalities and derived constraints} \label{sec:traces}

In light of the observations regarding the entanglement entropy and the smearing of spectral splits, one can attempt to formally relate the entropies $S(\omega)$ of individual neutrinos with their probabilities of being found in particular mass eigenstates (or equivalently, with the $z$-components of their polarization vectors). For this purpose, one might employ the Gibbs variational principle, which states that for any self-adjoint operator $Q$ such that $e^{-Q}$ is in the trace class, and for any $\gamma \geq 0$ with $\mathrm{Tr}\, \gamma = 1$, one can write down the inequality 
\begin{equation}
    \mathrm{Tr} (\gamma Q) + \mathrm{Tr} (\gamma \log \gamma) \geq - \log \mathrm{Tr} (e^{-Q}),
\end{equation}
with the equality satisfied if and only if $\gamma$ is given by $\gamma = e^{-Q}/\mathrm{Tr} (e^{-Q})$.

Taking $\gamma$ to be the reduced density matrix of a single neutrino $\rho_\omega^{\text{(red)}}$, as given by Eq.~\eqref{eq:reduced_density_mat}, one can obtain a sequence of inequalities with different choices of operators $Q$. Taking the trivial case $Q = \mathbb{I}$, i.e., the $2 \times 2$ identity matrix, one recovers the inequality $S(\omega) \leq \log(2)$, using the definition of $S(\omega)$ from Eq.~\eqref{eq:ent_form}. Taking $Q = \sigma_z$ instead, one obtains a constraint relation connecting the polarization vector component $P_{z,\omega}$ and the entanglement entropy $S(\omega)$:
%Taking $Q = J^z_\omega = \sigma_z/2$, one obtains a constraint relation connecting the polarization vector component $P_{z,\omega}$ and the entanglement entropy $S(\omega)$:\footnote{Henceforth, we drop the notation for  $\omega$-dependence in our inequalities for brevity.}
\begin{equation} \label{eq:gibbs1}
    P_{z,\omega} \ge  S(\omega) - \log \left(e + \frac{1}{e} \right).
\end{equation}

Similarly, for any component of $\vec{P}_\omega$ along a general direction, we can write the inequality 
\begin{equation} \label{eq:gibbsA}
\vec{A} \cdot \vec{P}_\omega + \log \big(2 \cosh |{\vec A}|\big) \ge S(\omega).
\end{equation}
where $\vec{A}$ is an arbitrary three-dimensional real vector. 
%(For example, note that if we set $\vec{A} = \vec{0}$, then we get the standard upper limit on entanglement entropy $S\leq\log(2)$.). 
One could also generalize Eq.~\eqref{eq:gibbs1} by taking $Q = n \sigma_z$, obtaining a series of such inequalities. Combining the constraints derived by taking $Q = \pm n \sigma_z$, one obtains the symmetric inequality
% \begin{equation} \label{eq:gibbs1}
%     P_z \ge S - \log \left(e + \frac{1}{e} \right),
% \end{equation}
% and combining it with the corresponding inequality for $Q = -\sigma_z$ gives the symmetric bound
% \begin{equation} \label{eq:gibbspm}
%     S \le \log \left(e + \frac{1}{e} \right) - |P_z|.
% \end{equation}
% Generalizing to $Q = \pm \, n \sigma_z$, one obtains
\begin{equation} \label{eq:gibbsn}
    S(\omega) \le \log \left(e^n + \frac{1}{e^n} \right) - n |P_{z,\omega}| \ \equiv \ S_n,
\end{equation}
where we have defined $S_n$ as the limiting expression on the right-hand side of the inequality. This can be turned around to yield a bound on the size of $|P_{z,\omega}|$ as a function of the entanglement entropy $S(\omega)$, namely,
\begin{equation} \label{eq:PzboundSn}
    |P_{z,\omega}| \le \frac1n \log \left(e^n + \frac{1}{e^n} \right) - \frac{S(\omega)}{n}.
\end{equation}

One may use such a bound to predict the smearing of the spectral split based on the degree of entanglement. For instance, from Fig.~\ref{fig:lomixspec16_256}, taking the particular neutrino at $\omega/\omega_0 = 13$ as an example, the entanglement entropy is approximately $S(\omega) \approx 0.5$, and therefore, taking Eq.~\eqref{eq:PzboundSn} with $n = 1$, one can derive the bound $P_{z,\omega} < \log (e + 1/e) - 0.5 \approx 0.6$, or equivalently, $P_{\nu_1}(\omega) = 1/2 \, (1+P_{z,\omega}) \leq 0.8$, which suggests that the spectral split is being smeared in the many-body case, compared to the mean-field result which has $P_{z,\omega} \approx 1$ at that frequency.
% Taking the limit $n \rightarrow \infty$ recovers the trivial result $|P_z| \leq 1$.
% The formula Michael and Baha discussed:
% \begin{equation*}
% S \le \frac{P_z}{2} + 0.81326 .
% \end{equation*}
% The constant is 
% \begin{equation*}
% \log \left( \sqrt{e} + \frac{1}{\sqrt{e}} \right) = 0.81326
% \end{equation*}

%The point: In this sense, the presence of entanglement entropy sets a bound on the size of the polarization vector, and in principle keeps it from reaching values of $\pm1$. One consequence of this bound is that the spectral split involves neutrinos with values of $\lim_{t\to\infty}|P_z(\omega)|<1$. 

However, it can be demonstrated that none of the bounds derived from the Gibbs variational principle are stronger than the bounds on $P_z$ (or other polarization vector components) imposed by the relation between entanglement entropy and the length of the polarization vector, $|\vec{P}|$, given by in Eqs.~\eqref{eq:ent_form} and~\eqref{eq:den_mat_eig} (see also Ref.~\cite{Cervia:2019res} for a closed form expression). 
% \begin{equation}
%     S = - \frac{1-P}{2} \log \left( \frac{1-P}{2} \right) -  \frac{1+P}{2}  \log \left(\frac{1+P}{2} \right).
%     \label{eq:concEntropy}
% \end{equation}
In fact, each of these constraints derived from Eq.~\eqref{eq:gibbsn} for various values of $n$ can be represented as tangents of the constraint in Eqs.~\eqref{eq:ent_form} and~\eqref{eq:den_mat_eig}. This relation is depicted in Fig.~\ref{fig:entpolconstr}. Even though the constraints based on the Gibbs variational principle are weaker than the one derived from Eqs.~\eqref{eq:ent_form} and~\eqref{eq:den_mat_eig}, they do nevertheless furnish straightforward, linear relations between $P_z$ and $S$ of individual neutrinos, which may be utilized as shown above.

\begin{figure}[htb]
    \begin{center}
        \includegraphics[width=0.48\textwidth]{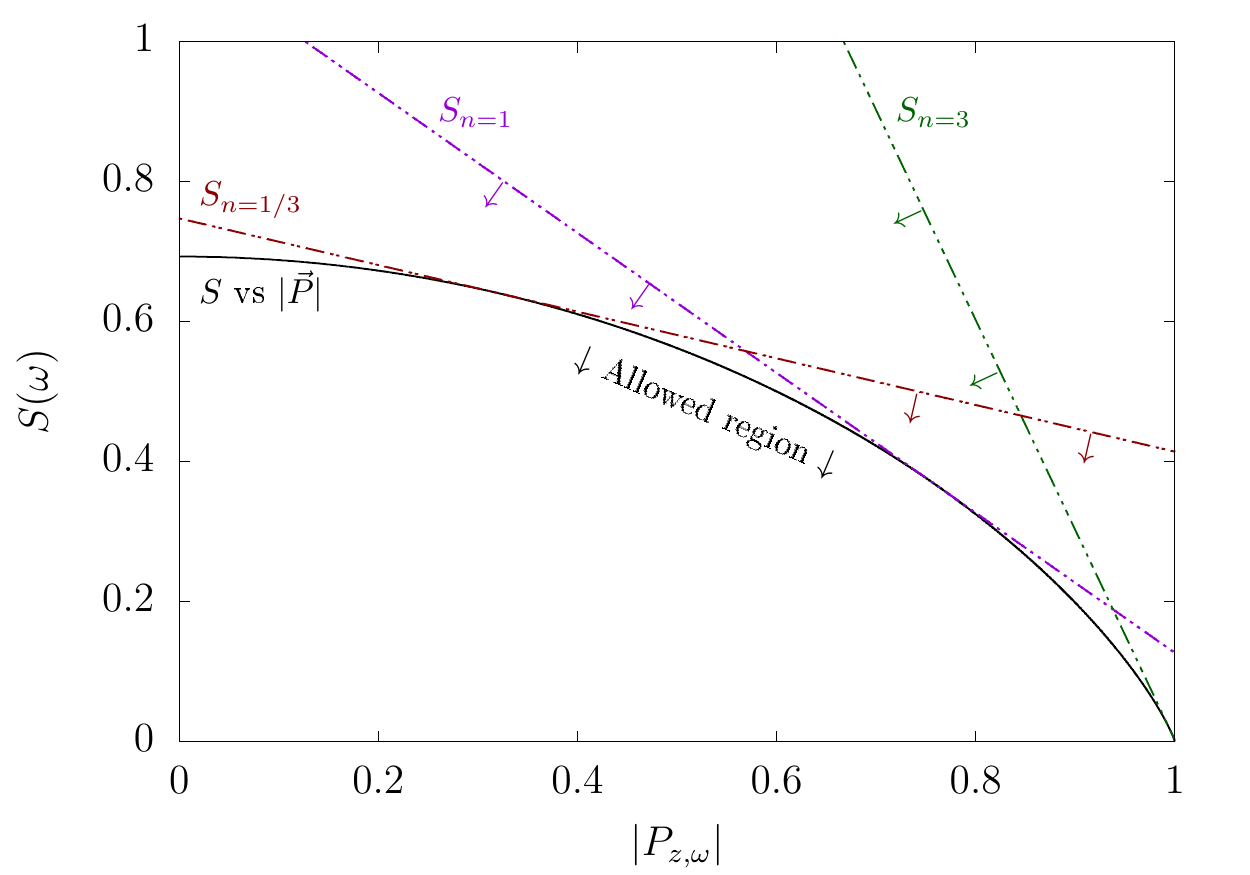}
    \end{center}
    \caption{A juxtaposition of the constraints derived from the Gibbs variational principle (Eq.~\eqref{eq:gibbsn}) and the constraint derived from the relation between entanglement entropy of individual neutrinos, $S(\omega)$, and the length of the neutrino polarization vectors, $|\vec{P}_\omega|$ (Eqs.~\eqref{eq:ent_form} and~\eqref{eq:den_mat_eig}). The dot-dashed straight lines are the limiting lines $S_n$ from Eq.~\eqref{eq:gibbsn}, for $n=1$ (purple), $n=3$ (green), and $n=1/3$ (dark red), respectively. The regions above and to the right of each of these lines in the $S(\omega)$ vs.~$|P_{z,\omega}|$ space are excluded. The solid (black) line represents the relation between $S(\omega)$ and $|\vec{P}_\omega|$. Since $|P_{z,\omega}| \leq |\vec{P}_\omega|$, this curve can also be considered as a limiting case of the permitted $P_{z,\omega}$ values as a function of the entanglement entropy. As is apparent, Eqs.~\eqref{eq:ent_form} and~\eqref{eq:den_mat_eig} furnish the strongest constraint on $P_{z,\omega}$ vs.~$S(\omega)$, whereas the other limiting lines derived from the Gibbs variational principle can be seen to be tangents of this curve.}
    \label{fig:entpolconstr}
\end{figure}

%\emph{Incorporation of Sec.~\ref{sec:traces} into explanation about spectral split behavior: }
From these relations, and as demonstrated using the simple example above, one can observe that the entanglement entropy growth tightens the bound on the size of the polarization vectors corresponding to the neutrinos most entangled with the rest of the ensemble. As a consequence, we can see that neutrinos closest to the spectral split, which we have found to have greater entanglement entropy, also have values $\lim_{t\to\infty}|P_{z,\omega}|<1$. In this sense, entanglement entropy of several neutrinos with frequencies around the split frequency result in a broadening of the split in many-body theory, even for calculations performed using the single-angle approximation. Interestingly, such broadening of the spectral split may also be observed in the mean field limit, with the inclusion of multi-angle effects~\cite{Dasgupta:2009mg} or non-standard $\nu$-$\nu$ interactions~\cite{Das:2017iuj}.

\hskip 0.3cm

\section{Conclusions} 
\label{sec:conclusions}

We showed that many-body calculations, in a single-angle approximation, predict a smeared spectral split---in contrast with mean-field calculations, which predict smearing only in certain multi-angle scenarios or with the addition of a non-standard self-interaction potential. Furthermore, we find that the width of this smeared spectral split is dependent upon the size of the mixing angle. Additionally, we note that growth in entanglement entropy is most substantial around the spectral split and note the role of entanglement in the smearing of the spectral split. 
%While we obtain the results summarized above, we demonstrated the relationship between the entanglement entropy and the third component of the polarization vector numerically as well as formally using the Gibbs' variational principle. 
Along with the results summarized above, we explain the smearing of the spectral split in terms of the relationship between the entanglement entropy and the third component of the polarization vector, which we establish numerically as well as formally using the Gibbs variational principle. 

The results presented in this article were established in calculations with up to 16 neutrinos. {As a result of being limited in terms of neutrino number, and due to the various physical assumptions used in this work (single-angle approximation, plane wave neutrinos, etc.), it remains to be seen whether the results obtained here could be considered to be representative of an actual core-collapse supernova environment. The goal of this study was to extend previous explorations of the potential effects of quantum entanglement on collective neutrino flavor evolution, using simplified numerical models. Looking forward, it is clearly desirable to increase the number of neutrinos treated in these calculations. Through this process, we intend to explore how the entanglement and the associated flavor phenomena scale with neutrino number would offer some clues about the large-$N$ limit of such systems (i.e., approaching the realistic number of neutrinos present in core-collapse supernovae and neutron-star mergers).} A more technical analysis comparing various pros and cons of several numerical approaches towards this goal is beyond the scope of this work, but this issue will be discussed in a future paper~\cite{Cerviaetal}.

Our results suggest that a full many-body calculation may not be necessary in all cases. At least in some cases, one can first run a mean-field calculation and obtain the split frequencies, and once they are determined, a many-body calculation may be run only for those neutrinos with energies near the split frequencies, keeping the mean-field results for other neutrinos. Such a hybrid approach could certainly cut down the computational time needed. One method to implement such a hybrid of entangled and non-entangled particles in the same ensemble will be presented in future work~\cite{Cerviaetal}. 

% \vskip 1.4cm

\begin{acknowledgments}
We thank S.~Coppersmith, C.~Johnson, and E.~Rrapaj 
%Y.~Pehlivan, E.~Rrapaj, and P.~Claeys 
for helpful conversations. This work was supported in part by the U.S. Department of Energy, Office of Science, Office of High Energy Physics, under Award No.~DE-SC0019465. 
% %and DE-FG02-03ER41272. 
It was also supported in part by the U.S. National Science Foundation
Grants No.~PHY-2020275 and PHY-2108339. The work of A.~V.~P. was supported in part by the NSF (Grant no. PHY-1630782)
and the Heising-Simons Foundation (2017-228), and in part by the U.S. Department of Energy under contract number DE-AC02-76SF00515.

%Tensor network calculations were performed using the TeNPy Library (version 0.8.4)~\cite{tenpy}.
\end{acknowledgments}

\bibliography{references}% Produces the bibliography via BibTeX.

\end{document}